\documentclass[11pt,a4paper]{article}

\usepackage{jheppub}
\pdfoutput=1
\usepackage[utf8]{inputenc}
\usepackage{float}
\usepackage{lmodern}

\usepackage{amsmath,amssymb,mhchem}
\usepackage{graphicx}
\usepackage{url}
\usepackage{hyperref}
\usepackage{color}
\usepackage{enumitem}
\usepackage{subfigure}
\usepackage[dvipsnames]{xcolor}
\hypersetup{allcolors=[rgb]{0.0 0.0 0.6},linkcolor=[rgb]{0.75 0.05 0.05}}
\usepackage{bm}
\usepackage{epsfig}
\usepackage{amsmath}
\usepackage{amssymb}
\usepackage{slashed}
\usepackage{color}
\usepackage{accents}
\usepackage{url}
\usepackage[dvipsnames]{xcolor}
\usepackage{cancel}
\usepackage[normalem]{ulem} 
\usepackage[colorinlistoftodos]{todonotes}
\usepackage{multirow}
\usepackage{natbib}
\usepackage{dsfont}
\usepackage{bbold}

\newcommand{\exclude}[1]{}

\newcommand{\pt}{\ensuremath{p_{\mathrm{T}}}}

\newcommand{\ptmiss}{\ensuremath{\pt^\text{miss}}}
\newcommand{\nhits}{\ensuremath{N_{\mathrm{hits}}}}

\def\vec{\mathaccent "017E\relax }

\newcommand{\ptvecmiss}{\ensuremath{{\vec p}_{\mathrm{T}}^{\kern1pt\text{miss}}}}

\begin{document}

\preprint{IFIC/22-xx}
\preprint{FERMILAB-PUB-22-787-CMS-PPD-QIS}

\author[a,b]{Giovanna Cottin,}
\emailAdd{giovanna.cottin@uai.cl}
\affiliation[a]{Departamento  de  Ciencias,  Facultad  de  Artes  Liberales, \\ Universidad  Adolfo  Ib\'a\~nez,  Diagonal  Las  Torres  2640,  Santiago,  Chile}
\affiliation[b]{Millennium Institute for Subatomic Physics at the High Energy Frontier (SAPHIR), Fern\'andez Concha 700, Santiago, Chile}

\author[c,b]{Juan Carlos Helo,}
\emailAdd{jchelo@userena.cl}
\affiliation[c]{Departamento de F\'{i}sica,
Facultad de Ciencias, Universidad de La Serena,
Avenida Cisternas 1200, La Serena, Chile}

\author[d]{Martin Hirsch}
\emailAdd{mahirsch@ific.uv.es}
\affiliation[d]{AHEP Group, Instituto de F\'{\i}sica Corpuscular --
    CSIC/Universitat de Val{\`e}ncia, Apartado 22085,
  E--46071 Val{\`e}ncia, Spain}

\author[e]{Cristi\'an Pe\~na,} 
\emailAdd{cmorgoth@fnal.gov}
\affiliation[e]{Fermi National Accelerator Laboratory, Batavia, IL 60510, U.S.A.}

\author[f]{Christina Wang,}
\emailAdd{christina.wang@caltech.edu}
\affiliation[f]{California Institute of Technology, Pasadena, CA 91125, U.S.A.}

\author[e,f]{Si Xie,} 
\emailAdd{sixie@caltech.edu}

\date{\today}

\title{Long-lived heavy neutral leptons with a displaced shower signature at CMS}

\date{\today}

\abstract{We study the LHC discovery potential in the search for heavy neutral leptons (HNL) with a new signature: a displaced shower in the CMS muon detector, giving rise to a large cluster of hits forming a displaced shower. A new Delphes module is used to model the CMS detector response for such displaced decays. We reinterpret a dedicated CMS search for neutral long-lived particles decaying in the CMS muon endcap detectors for the minimal HNL scenario. We demonstrate that this new strategy is particularly sensitive to active-sterile mixings with $\tau$ leptons, due to hadronic $\tau$ decays. HNL  masses between $\sim 1 - 6$ GeV can be accessed for mixings as low as $|V_{\tau N}|^{2}\sim 10^{-7}$, probing unique regions of parameter space in the $\tau$ sector.  
}

\maketitle
{
  \hypersetup{linkcolor=black}
  \tableofcontents
}


\section{Introduction}
\label{sec:intro}
Long-lived particles (LLPs) are highly motivated on theoretical and experimental grounds~\cite{Curtin:2018mvb,Alimena:2019zri}. 
In the context of theories that address the mechanism for the generation of light neutrino masses in the Standard Model (SM), a new sterile or right-handed neutrino that can be long-lived is predicted in several models. This new long-lived sterile neutrino will decay at displaced locations inside the Large Hadron Collider (LHC) detectors thus creating striking experimental signatures with great discovery potential.
In the minimal type-I seesaw framework, the new right-handed fermion singlet, hereafter referred to as heavy neutral lepton (HNL), mixes with the SM neutrinos~\cite{Minkowski:1977sc,Yanagida:1979as,GellMann:1980vs,Mohapatra:1979ia,Schechter:1980gr}.
For sufficiently low mixing and HNL masses below the electroweak scale, the HNL can be long-lived~\cite{Helo:2013esa}.

The ATLAS collaboration has recently searched for a long-lived HNL produced in the decays of $W$ bosons at 13 TeV decaying to $l=e,\mu$~\cite{ATLASnewHNL,Aad:2019kiz}, excluding HNL masses in the range $\sim 4.5-10$ GeV for mixings as low as $ |V_{lN}|^{2}\approx 10^{-6}$. 
CMS has also recently searched for HNLs~\cite{CMS:2022fut}, where two leptons from the HNL decay are identified and form a displaced secondary vertex inside the CMS silicon tracker. 
This search excludes HNLs in the mass range $1-20$ GeV for mixings as low as $ |V_{lN}|^{2}\approx 10^{-7}$, with $l=e,\mu$. 
As far as we know, there are no dedicated HNL searches at the LHC focusing on mixings in the tau sector. For prompt phenomenological LHC prospects, see for instance~\cite{Cheung:2020buy}.

In recent years, several phenomenological studies aim to access different regions in the mixing-mass HNL plane, for mixing with different flavours. 
These include  lepton-jets~\cite{Izaguirre:2015pga}, displaced vertices in inner-trackers~\cite{Beltran:2021hpq,Cottin:2018nms,Abada:2018sfh}, displaced vertices in muon chambers~\cite{Bondarenko:2019tss,Drewes:2019fou,Boiarska:2019jcw} or displaced leptons~\cite{Liu:2019ayx}. 
These current searches use triggers containing either one or two prompt leptons, and are most efficient in constraining mixings in the electron and muon sectors.

In this work, we study the sensitivity to long-lived HNLs which produce a displaced shower signature when decaying inside the CMS muon system. Previous proposals for detecting long-lived HNLs involving the muon chambers relied on tracker-based information to reconstruct a displaced vertex signal~\cite{Bondarenko:2019tss,Drewes:2019fou,Boiarska:2019jcw}. These have the advantage of constraining larger HNL lifetimes (and therefore lower masses) as opposed to displaced vertex searches in the inner trackers of LHC detectors~\cite{Cottin:2018nms,Beltran:2021hpq}. Here we propose a different and complementary search strategy, where the muon spectrometer is re-purposed as a calorimeter.

A novel CMS search~\cite{CMS:2021juv}, originally interpreted in the context of a SM Higgs boson decaying to long-lived scalars, uses the CMS muon detector as a sampling calorimeter to identify showers produced by LLPs decaying to final states including hadrons, taus, electrons, or photons. The analysis strategy originally considers a trigger on missing transverse momenta, and uses the unique features of the CMS muon detector to identify high-multiplicity hit clusters that form a displaced shower. LLPs decays, including those of HNLs, in the CMS muon system will induce hadronic and electromagnetic showers. 

In the case of displaced long-lived HNLs decays, the shower in the CMS muon system will emerge from the subsequent decay of the off-shell $W$ as well as from the associated displaced electron and $\tau$ in the electron and $\tau$ sectors, respectively. 
The higher cluster reconstruction efficiency for hadronic showers leads to an enhanced sensitivity when the HNL decays to a hadronically decaying $\tau$ lepton. 
Although we expect the displaced shower signature of the CMS muon system to have sensitivity to HNLs in all three lepton sectors, in this study, we will not consider the scenario with mixing in the muon sector due to insufficient information available to estimate the detector response for displaced muons and to obtain an accurate signal yield (see Sec.~\ref{sec:results} for details).

Furthermore, the large amount of shielding provided by the steel interleaved in the muon detectors provide excellent rejection of hadronic backgrounds, which is a driving factor in the sensitivity for displaced HNLs decaying with a displaced shower.   

Another advantage of this strategy compared to past displaced HNL search proposals in the $\tau$ sector~\cite{Cottin:2018nms,Beltran:2021hpq}, is that it has no limitation to access HNL masses below 5~GeV. 
In past displaced HNL search proposals~\cite{Cottin:2018nms,Beltran:2021hpq}, a requirement on the invariant mass of the displaced vertex to be above 5~GeV was needed to suppress backgrounds from $B-$ mesons. 
In this analysis, the large amount of shielding in the CMS muon detector allows for large background suppression, which is particularly important for signatures with a {\it{single}} displaced LLP exhibited by HNL signal scenario. 
Motivated by the critical need for dedicated displaced object triggers, we also recast the CMS analysis considering a new dedicated displaced trigger. 
A new L1 trigger proposed for Run 3 of the LHC~\cite{LLPCMSL1Trigger,Alimena:2021mdu} would allow to trigger directly on the HNL signature, raising the search sensitivity by several orders of magnitude in the HNL mixing-mass plane.

The paper is organized as follows. 
In section~\ref{sec:HNLtheory} we detail and motivate HNL models coupling to only one generation of SM charged leptons. In section~\ref{sec:AS} we describe our analysis strategy, the reinterpretation procedure employed, and the development and usage of a new Delphes module for displaced showers. 
Exclusion limits for the recasted CMS search, as well as estimates with the new displaced trigger, are presented in section~\ref{sec:results} for the minimal HNL model. We conclude in section~\ref{sec:conclusion}.

\section{HNL interactions and neutrino mass models}
\label{sec:HNLtheory}

We begin by defining the {\em minimal HNL model} and then discuss its relation to two basic seesaw models: The classical seesaw type-I \cite{Minkowski:1977sc,  Yanagida:1979as,Mohapatra:1979ia,GellMann:1980vs,Schechter:1980gr} and the inverse seesaw \cite{Mohapatra:1986bd}.

An HNL is defined by its charged and neutral current interactions with
standard model leptons:
\begin{eqnarray}\label{CC-NC}
{\cal L}_{\rm int} &=& \frac{g}{\sqrt{2}}\, 
 V_{\alpha N_j}\ \bar l_\alpha \gamma^{\mu} P_L N_{j} W^-_{L \mu} 
+\frac{g}{2 \cos\theta_W}\ \sum_{\alpha, i, j}V^{L}_{\alpha i} 
V_{\alpha N_j}^*  \overline{N_{j}} \gamma^{\mu} P_L \nu_{i} Z_{\mu} 
+ {\rm h.c.}
\end{eqnarray}
Here, $V_{\alpha N_j}$ are free parameters, parametrizing the mixing
angle of $N_j$ and, in principle, one can add $j=1, \cdots, n$ HNLs to
the SM. In searches, one typically assumes there is only one HNL with a
mass in the kinematically accessible region. Note, $V^{L}_{\alpha i}$
is the mixing among light neutrinos.

To ${\cal L}_{\rm int}$ one has to add a mass term. This mass term 
could be either of Dirac or Majorana type. For Dirac HNLs, only 
lepton number conserving decays (LNC) are possible, whereas Majorana 
HNLs can have both LNC and lepton number violating (LNV) decays. Thus, 
for the same values of $m_N$ and $V_{\alpha N_j}$, the decay width of 
a Majorana neutrino is twice that of a Dirac neutrino. For definiteness, 
in the numerical part of this work we use Majorana HNLs. 

Eq. (\ref{CC-NC}) gives the interaction Lagrangian for an HNL at
$d=4$.  An HNL could also have additional non-renormalizable
interactions, see for example
\cite{delAguila:2008ir,Aparici:2009fh,Liao:2016qyd}. However, we will
disregard this possibility and define the minimal HNL model as the one
based on eq.  (\ref{CC-NC}).

The study of HNLs is usually motivated by the observed neutrino 
masses, see for example \cite{deSalas:2020pgw} for a recent overview 
on the status of neutrino data. The minimal HNL model, on the 
other hand, takes the $V_{\alpha N_j}$ for $\alpha=e,\mu,\tau$ as free 
parameters and does not explain light neutrino masses. To make 
contact with neutrino data one needs to connect the HNL with some 
theoretical neutrino mass model.

The simplest possible model is the type-I seesaw.  In seesaw type-I
one adds three right-handed neutrinos
\footnote{Current data allows for one active neutrino to be
  massless. Thus, in principle, only two right-handed neutrinos are
  necessary to explain the data.}  to the standard model field
content. The model generates the mass matrix for the six neutral states:
\begin{equation}
\mathcal M_{\rm type-I}=\left(\begin{array}{c c} 
0 & m_D^T \\ 
m_D & M_R
\end{array}\right) \, .
\label{eq:typeImatrix}
\end{equation}
Here $m_D$ is Dirac mass matrix, while $M_R$ is the Majorana mass
matrix for the right-handed singlets. In seesaw type-I one can always
choose to work in the basis where the latter is diagonal, 
${\hat M_R}$.  After diagonalization of eq. (\ref{eq:typeImatrix}) the
light neutrino masses and the mixing between the light (and mostly
active) and heavy (mostly sterile) neutrinos is given by
\begin{eqnarray}
m_{\nu} &=& - m_D^T \cdot M_R^{-1}\cdot m_D + \cdots
\label{eq:typeIMass}
\\
V_{H-L} &=& m_D^T \cdot M_R^{-1} + \cdots ,
\label{eq:typeIMix}
\end{eqnarray}
where the dots represent higher order terms. 
Note that the matrix elements of $V_{H-L}$ correspond to the mixing angle parameters  $V_{\alpha N_j}$, in eq. (\ref{CC-NC}), but we use a different symbol to distinguish it from the 
``model-independent'' HNL setup. For the seesaw type-I, one 
can find a simple reparametrization of the Dirac mass matrix 
\cite{Casas:2001sr}:
\begin{equation}
m_D = i \sqrt{{\hat M_R}}\cdot {\cal R}\cdot \sqrt{{\hat m}_{\nu}}
\cdot U_{\nu}^{\dagger}.
\label{eq:CI}
\end{equation}
Eq. (\ref{eq:CI}) guarantees that the seesaw parameters chosen always
fit the input neutrino data. Here, $U_{\nu}$ is
the mixing matrix observed in oscillation experiments, ${\hat
  m}_{\nu}$ and ${\hat M_R}$ are the eigenvalues for the light and
heavy neutrinos respectively and ${\cal R}$ is an orthogonal matrix of
three complex angles. The entries in ${\cal R}_i$ can be written in
terms of $s_i\equiv \sin(z_i)$, with $z_i=\kappa_i \times e^{2 i \pi
  \xi_i}$ \cite{Anamiati:2016uxp}. It is straightforward to show that
for all $s_i=0$, the matrix $V_{H-L}$ is given by:
\begin{equation}
(V_{H-L})_{ij} = (U_{\nu}^{*})_{ij} \sqrt{\frac{{\hat m}_{\nu,i}}{M_{R,i}}}.
\label{eq:typeIMixSim}
\end{equation}
Thus, one expects that the mixing $V_{H-L}$ is suppressed by light
neutrino masses in type-I seesaw. Also, in this limit the branching
ratios of the heavy singlets to the different SM generations are
related to measured neutrino angles.  However, for complex ${\cal R}$
one can find larger values of $V_{H-L}$, if one allows the different
contributions in eq. (\ref{eq:typeIMass}) to nearly cancel against
each other. Note that in this fine-tuned part of parameter space,
eq. (\ref{eq:CI}) is no longer valid, since 1-loop corrections to the
seesaw become more important than the tree-level itself, see the
discussion in \cite{Cordero-Carrion:2019qtu}.  While in this
``cancellation region'' one can have mixings large enough to be
experimentally testable, only very few and fine-tuned points exist 
in this particular part of parameter space of the seesaw \cite{Feng:2022inv}, 
for which the right-handed neutrinos can decay to a single SM lepton 
generation.  This conclusion, however, is valid strictly only for type-I seesaw.

Very different expectations for $V_{H-L}^2$ and $N_i$ branching ratios
are obtained for the inverse seesaw mechanism. (We concentrate on
inverse seesaw here, but a similar discussion could be presented for the
{\em linear} seesaw \cite{Akhmedov:1995ip,Akhmedov:1995vm}.) In
inverse seesaw \cite{Mohapatra:1986bd}, three additional singlets,
denoted $S$, are added and the ($9,9$) mass matrix is given by:
\begin{equation}
\mathcal M_{\rm ISS}=\left(\begin{array}{c c c} 
0 & m_D^T & 0 \\ 
m_D & 0 & M_R \\ 
0 & M_R^T & \mu 
\end{array}\right) \, .
\label{eq:ISSmatrix}
\end{equation}
Note that, in the limit $\mu\equiv 0$ the three active neutrinos are
massless, i.e. lepton number is conserved.  Thus, a small value of
$\mu$ is technically natural. In this limit the 6 heavy
states form three Dirac pairs with masses $M_{R_i}$.  For $\mu \ll m_D
\ll M_R$, the mass matrix for the lightest three states, the 
masses of the heavy states and the mixing
to the heavy neutrinos are given as:
\begin{eqnarray}
(m_\nu)_{\rm ISS} & = & m_D^T \cdot {M_R^T}^{-1} \cdot \mu \cdot M_R^{-1} \cdot m_D 
+ \cdots
\label{eq:numassISS}
\\
M_{\pm} & \simeq &  
\Big( {M}_R + \left\{m_D . m_D^T,M_R^{-1}\right\} \Big) 
\pm \frac{1}{2}{\mu}
\label{eq:numassISSH}
\\
V_{H-L} &=& \frac{1}{\sqrt{2}} m_D^T \cdot M_R^{-1} + \cdots \simeq
\sqrt{\frac{{\hat m}_{\nu}}{\mu}}
\label{eq:MixISS}
\end{eqnarray}
Here $\{ a,b\}$ is the anti-commutator of $a$ and $b$.  The heavy
states thus form ``pseudo-Dirac'' pairs, splitted by the small
parameter $\mu$. In the limit $\Gamma \ll \mu$, where $\Gamma$ is the
total decay width of the heavy state, the singlets behave as Majorana
particles, while for the opposite limit $\mu \ll \Gamma$, the decays
are all Dirac-like \cite{Anamiati:2016uxp}.  Heavy-light mixing in
inverse seesaw is given by the same ratio of $m_D$ and $M_R$ as for
seesaw type-I, but the relation of $V_{H-L}$ to light neutrino masses
is changed, thus the second relation in eq. (\ref{eq:MixISS})
above. Clearly, the naive expectation is that for an inverse seesaw
model, the mixing $V_{H-L}$ is {\em enhanced by a factor ${\hat
    M_R}\cdot \mu^{-1}$} relative to the seesaw type-I.

One can formulate a parametrization of $m_D$ in terms of neutrino
oscillation parameters, $M_R$, $\mu$ and ${\cal R}$
\cite{Cordero-Carrion:2019qtu} in the same spirit as the Casas-Ibarra
parametrization for the type-I seesaw \cite{Casas:2001sr}. For ${\cal
  R}=\mathbb{1}$ one obtains the second equation in
eq. (\ref{eq:MixISS}) above. The larger number of free parameters in
the inverse seesaw, however, allows not only to easily find parameter
space with much larger $V_{H-L}$ than for the seesaw type-I, it also
offers the possibility to break the relation $V_{H-L} \propto
U_{\nu}^{*}$, shown in  eq. (\ref{eq:typeIMixSim}). The simplest possibility to do so, is to choose both 
$m_D$ and $M_R$ diagonal.\ In this case, according to eq. (\ref{eq:MixISS}) $V_{H-L}$ will be diagonal, and therefore each of the three (pairs of) heavy singlets will decay to only one generation of SM leptons. Even in this case, the neutrino data can be correctly fitted easily as can be seen in the following expression derived from eq. (\ref{eq:numassISS}): 
\begin{equation}\label{eq:Fitmu}
\mu = M_R^T\cdot (m_D^T)^{-1} \cdot U_{\nu}^{*}\cdot {\hat m_{\nu}}
\cdot U_{\nu}^{\dagger} m_D^{-1} \cdot M_R
\end{equation}
In subsequent sections we will denote  these theoretical scenarios as electron-type, muon-type, and tau-type HNL.

The above discussion, while by far not covering all theoretical
possibilities, serves to show that from the point of view of neutrino
model building, larger HNL mixing is expected in the inverse seesaw
model. Moreover, discovering a HNL with ``large'' mixing, but coupling
to only one generation of SM charged leptons would be a strong hint
that the underlying neutrino mass model is not the simplest type-I
seesaw. In the numerical part of this work, we will, however, use the
minimal HNL model, treating $V_{\alpha N}$ simply as free parameters.

\section{Analysis strategy and simulation in Delphes}\label{sec:AS}

We consider a long-lived HNL that couples to the SM leptons via a small mixing in the electroweak currents, as detailed in the previous section.  The HNL ($N$) is produced at the LHC via $W$ bosons decaying leptonically: $W^{\pm}\rightarrow N l^{\pm}$, with $l=e, \mu$ or $\tau$. $N$ decays via charged and neutral currents, $N\rightarrow l^{\pm}q\bar{q}$, $N\rightarrow l'^{\mp}l^{\pm}\nu_{l}$,
and $N\rightarrow \nu_{l}q\bar{q}$ \cite{Helo:2013esa}. The relevant parameters are the HNL mass, $m_{N}$, and active-sterile neutrino mixing, $|V_{\alpha N}|^{2}$.

We generate HNL events using MadGraph5~\cite{Alwall:2011uj,Alwall:2014hca} and use Pythia 8~\cite{Sjostrand:2014zea} for parton showering, hadronization, and the HNL decay.
We use Delphes~\cite{delphes} and the associated CMS detector configuration card to simulate the detector response, along with a dedicated new module to simulate the response of the CMS muon detector for the HNL decay~\cite{delphes_pr}. 
Finally, we apply the selection requirements used by the CMS search for neutral LLPs in the endcap muon detector~\cite{CMS:2021juv}, calculate the expected HNL signal event yields, and compute limits for the minimal HNL model recasting the results of the CMS search. 

\subsection{Event generation and selections}

We use the FeynRules implementation for HNLs of Ref.~\cite{Degrande:2016aje} to generate events in Madgraph5 for $pp\rightarrow W$, with up to two jets and $W^{\pm}\rightarrow N l^{\pm}$. Samples with different jet multiplicities are merged according to the MLM prescription~\cite{Mangano:2006rw}. 
We apply generator-level cuts on (boosted) HNL kinematics, $\pt\geq$ 100~GeV and $0.5<|\eta|<$3, in MadGraph5 to increase the statistics in the phase space regions selected by the CMS analysis. 
At least 100,000 events are generated per mixing-mass point to maintain the statistical uncertainty for the predicted signal yield below $20\%$.

The leading order (LO) $W$ boson production cross section and the shape of the $W$ $\pt$ spectrum are  corrected to the best known theoretical prediction at next-to-next-to-leading order (NNLO)~\cite{dyturbo}. 
The NNLO correction yields a 30\% increase to the total $W$ boson production cross section and the $W$ boson $\pt$ spectrum correction increases the signal yield prediction by a factor of two.

\subsection{Delphes detector simulation}

\label{delphesModule}
The response of the CMS detector is simulated using Delphes~\cite{delphes}. The simulation uses the CMS detector configuration card~\cite{Mertens:2015kba}, producing a set of standard particle flow (PF) candidates. 
The simulation of the clusters of hits in the CMS cathode strip chamber (CSC) of the endcap muon detector is performed using a dedicated Delphes module developed~\cite{delphes_pr} using the parameterized detector response functions provided in the HEPData entry~\cite{hepdata.104408.v2}, associated with the CMS search result in Ref.~\cite{CMS:2021juv}. 
 In the CMS search, the CSC cluster is defined by grouping high-density CSC hit regions with a minimum of 50 hits. 
The number of hits comprising each cluster is defined as $\nhits$.
Experimentally, we expect clusters with $\nhits$ above a few hundred for signal, and clusters with a steeply falling distribution of $\nhits$ for background. 
More details about the CSC clusters are found in Ref~\cite{CMS:2021juv}.

The simulation of cluster-level selection efficiencies are divided into three components. The first component is the cluster efficiency, which includes the efficiencies of the cluster reconstruction, muon veto, active veto, time spread, and $\nhits$ requirements as used in the CMS search~\cite{CMS:2021juv}.
The cluster efficiencies are provided as a function of the LLP decay position in the CSC detector, and the associated electromagnetic and hadronic energies of its decay products.
Electromagnetic energy of the LLP is defined as the sum of the energies of any electrons or photons in the LLP decay chain. 
The hadronic energy of the LLP is defined as the sum of the energies of any other particles except for muons, neutrinos, or other BSM weakly interacting particles. 
A dedicated \textsc{CscClusterEfficiency} module in Delphes encoding the parameterized function was implemented. 
The second component is the cluster identification efficiency. 
The \textsc{CscClusterID} module following the code function provided by the CMS HEPData entry was implemented. 
The third component is model dependent and includes the cluster time requirement, the jet veto, and the $\Delta\phi$ requirement. 
The value of these 3 observables are calculated using generator-level information and these requirements are imposed later in the analysis workflow to predict the signal event yield. 
The cluster time is determined by calculating the LLP travel time from the production point to the decay vertex in the laboratory frame. 
The jet veto is implemented by requiring no jets with $\pt>$10 GeV and $\Delta R < 0.4$ between the jet and the LLP. 
Finally, $| \left. \Delta\phi (\mathrm{cluster,\ptvecmiss}) \right. |$ is defined by calculating the azimuthal difference between the LLP momentum and missing transverse momentum ($\ptvecmiss$), simulated using the standard Delphes modules.

\begin{figure}[ht]
    \centering
    \includegraphics[width=0.6\textwidth]{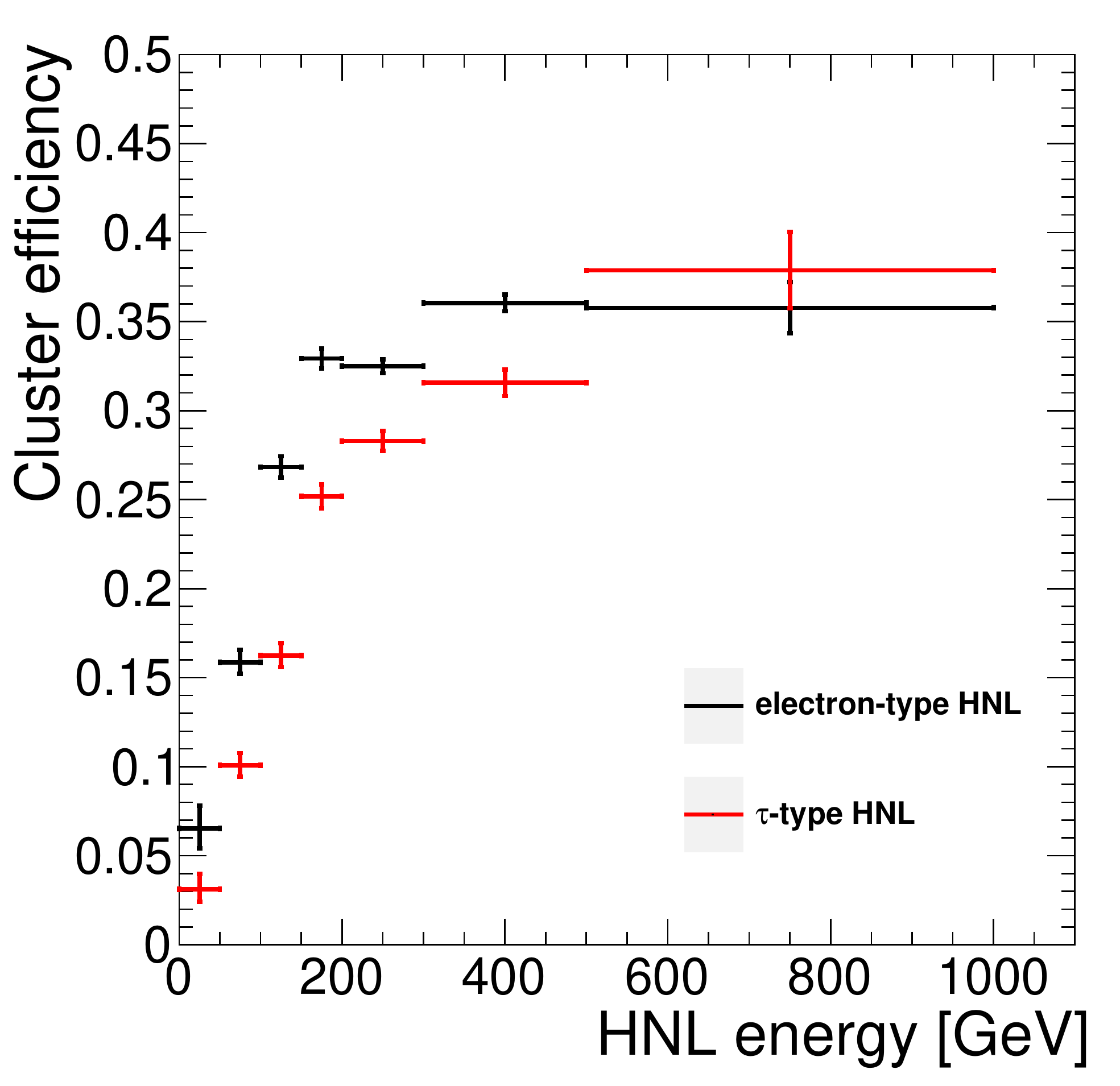}
    \caption{The reconstruction efficiencies of CSC cluster estimated by the \textsc{CSCClusterEfficiency} module as a function of the HNL energy are shown for electron- and $\tau$-type HNL decays. The HNLs have masses of 3~GeV and $c\tau$ of 1~\text{m}.}
    \label{fig:cluster_eff}
\end{figure}

The reconstruction efficiencies of CSC cluster passing the \textsc{CSCClusterEfficiency} module for HNLs decaying in region  with largest acceptance (region B), as defined in the HEPData entry~\cite{hepdata.104408.v2}, as a function of the HNL energy is shown in Figure~\ref{fig:cluster_eff}, for electron- and $\tau$-type HNL.
The efficiency increases as a function of the HNL energy and reaches a plateau at energies above 400~GeV. 
The $\tau$-type HNL efficiency is slightly lower because some fraction of the HNL energy is invisible due to $\tau$-lepton decays to neutrinos. 
However, given the same HNL visible energy, the $\tau$-type HNL would have higher cluster efficiency compared to electron-type HNL, due to a larger hadronic energy fraction.

\subsection{Reinterpretation and proposed strategies for high-luminosity LHC}\label{sec:reinterpretation}

We reinterpret the CMS Run 2 search result in Ref.~\cite{CMS:2021juv}, and also project the results to the integrated luminosity of the high luminosity LHC (HL-LHC)~\cite{Apollinari:2015wtw} considering two different search strategy proposals.

First, we perform a straight recasting of the Run 2 result, using the exact same selection as the published CMS result. 
We apply all the selection cuts, following the instructions provided in the Resources section in the HEPData entry~\cite{hepdata.104408.v2} as follows:
\begin{itemize}
    \item $|\ptvecmiss|\geq$ 200 GeV
    \item 1 CSC cluster that passes the \textsc{CSCClusterEfficiency} and \textsc{CSCClusterID} module
    \item $|\left. \Delta\phi (\mathrm{cluster,\ptvecmiss}) \right.| <$ 0.75
  
    \item CSC cluster time between -5 and 12.5 ns
    \item CSC cluster is not matched to jets with $\pt \geq$ 10 GeV
\end{itemize}
To recast the CMS Run 2 search result, we count the number of signal events passing the search selection requirements and predict the expected yield. A background prediction of 2$\pm$1 and the observed event yield in data, 3, are taken from the CMS~\cite{CMS:2021juv} result directly. 

To inform future experimental studies, we project the sensitivity of this analysis to future datasets, including to the end of Run 3 and to the end of the HL-LHC, where new dedicated triggers targeting this displaced signature would be available~\cite{LLPCMSL1Trigger}.
For future projections, we consider two different search strategies.

For search strategy 1, we maintain the use of the high $\ptmiss$ trigger, but apply a tighter $\nhits$ requirement in order to suppress the expected background close to zero, for an integrated luminosity of 3~ab$^{-1}$, representing the dataset expected for the end of the HL-LHC.
We fit the $\nhits$ distributions, provided in the additional materials from the CMS search result in Ref.~\cite{CMS:2021juv}, to an exponential function and extrapolate the expected background yield to larger values of $\nhits$. We find that requiring $\nhits > 210$ would suppress the expected background yield to 0.2 for an integrated luminosity of 3~ab$^{-1}$.
Based on the signal $\nhits$ distributions in the same plot, we find that increasing the $\nhits$ threshold from 130 to 210 would reject an additional 20\% of signal events, so we decrease the predicted signal yield accordingly.

For search strategy 2, we consider the scenario in which a dedicated Level-1 and High Level Trigger targeting this displaced signature is enabled and project the sensitivity for Run 3 and HL-LHC, respectively.
Because of the capability of the dedicated displaced trigger, we no longer need to impose a very high threshold requirement for $\ptmiss$ that was required by the existing CMS search~\cite{CMS:2021juv}.
By reducing the $\ptmiss$ requirement from 200 GeV to 50 GeV we can increase the signal acceptance by three orders of magnitude.
The rate of the main background, $W+$jets production, consequently increases by the same factor.
We suppress the background to near negligible levels again by increasing the $\nhits$ to 290, resulting in a background yield of 0.2 for a dataset with an integrated luminosity of 300 fb $^{-1}$, representing the dataset expected for the end of Run 3.
We find that the signal yield increase due to the new trigger significantly offsets the 40\% signal yield decrease due to the $\nhits >$ 290 with respect to the nominal $\nhits$ threshold at 130.
We also consider strategy 2 applied to the HL-LHC dataset with an integrated luminosity of 3~ab$^{-1}$, which requires to increase the $\nhits$ threshold to 370 in order to suppress the background yield to 0.2.
Accordingly, the signal yield reduces by 50\% with respect to the signal yield obtained for the nominal $\nhits$ threshold at 130.

The average number of pileup interactions will increase to 200 for the HL-LHC~\cite{Apollinari:2015wtw}, resulting in worse object reconstruction efficiency, degraded $\ptmiss$ resolution, and a larger number of spurious jets. 
We assume that improved pileup mitigation algorithms and upgraded detectors, including the MIP Timing Detector (MTD)~\cite{CMS:2667167}, will be able to mitigate the impact of the additional pileup on the object reconstruction and resolution.
However, we find that the probability for a cluster produced by an HNL decay in the CMS endcap muon detector to be accidentally matched to and vetoed by a pileup jet with $\pt>10$ GeV is 20\% larger. 
Therefore, we decrease the cluster reconstruction efficiency by 20\% for the HL-LHC scenarios explored in strategies 1~and~2. 
For all cases, we propagate a 20\% signal systematic uncertainty, accounting for the uncertainty in the efficiency of the cluster vetos and the $\ptmiss$ requirement.

Given the signal and background yield estimate, we evaluate the 95\% confidence level (CL) limits using the ``modified frequentist criterion" CL$_\mathrm{s}$~\cite{Read_2002} for each point in the parameter space.
The recasting procedure has been validated against the CMS exclusion limits, by recasting the twin Higgs model used in the CMS paper, as detailed in Appendix~\ref{sec:validation}.
The results of the recast of the Run 2 analysis, as well as the projections with strategies 1 and 2 mentioned above, are given in what follows.

\section{Results} \label{sec:results}  

Based on the search strategies described in the previous section, we have estimated the experimental sensitivity of the CMS muon system on HNLs. For simplicity we have considered a minimal HNL model that assumes there is only one  HNL with mass in the kinematic region of interest and that the mass of the HNL ($m_N $) and their mixing with the active neutrinos ($V_{lN}$) are free parameters.

We estimate the experimental sensitivity to the HNL minimal scenario in the $|V_{\alpha N}|^2$ vs $m_N$ plane for $\alpha= e, \tau$. 
The scenario of HNL mixing with muons is not considered due to insufficient information provided for the muon detector response for displaced muons produced in the muon detector volume to estimate the signal yield accurately. In the CMS result~\cite{CMS:2021juv}, a muon veto is implemented to reject clusters that are geometrically matched to muons, to reject clusters originating from muon bremsstrahlung. A displaced muon from the HNL decaying in the muon detector could be reconstructed, thus vetoing the signal cluster shower. However, the reconstruction efficiency of displaced muon in the muon detector is not known, so we cannot accurately estimate the signal efficiency of the muon veto for HNLs that mix with the muon sector. We leave the consideration of the scenario with mixing in the muon sector for a future study, once the necessary muon veto efficiencies for displaced muons have been provided by the CMS Collaboration. On the other hand, only a few percent of signal events in electron and $\tau$ sector contain displaced muons in the final state passing the muon veto threshold of $\pt>20$ GeV, so the impact on the sensitivity is negligible and is propagated as a source of signal systematic uncertainty.

\begin{figure}[ht]
    \centering
    \includegraphics[width=0.49\textwidth]{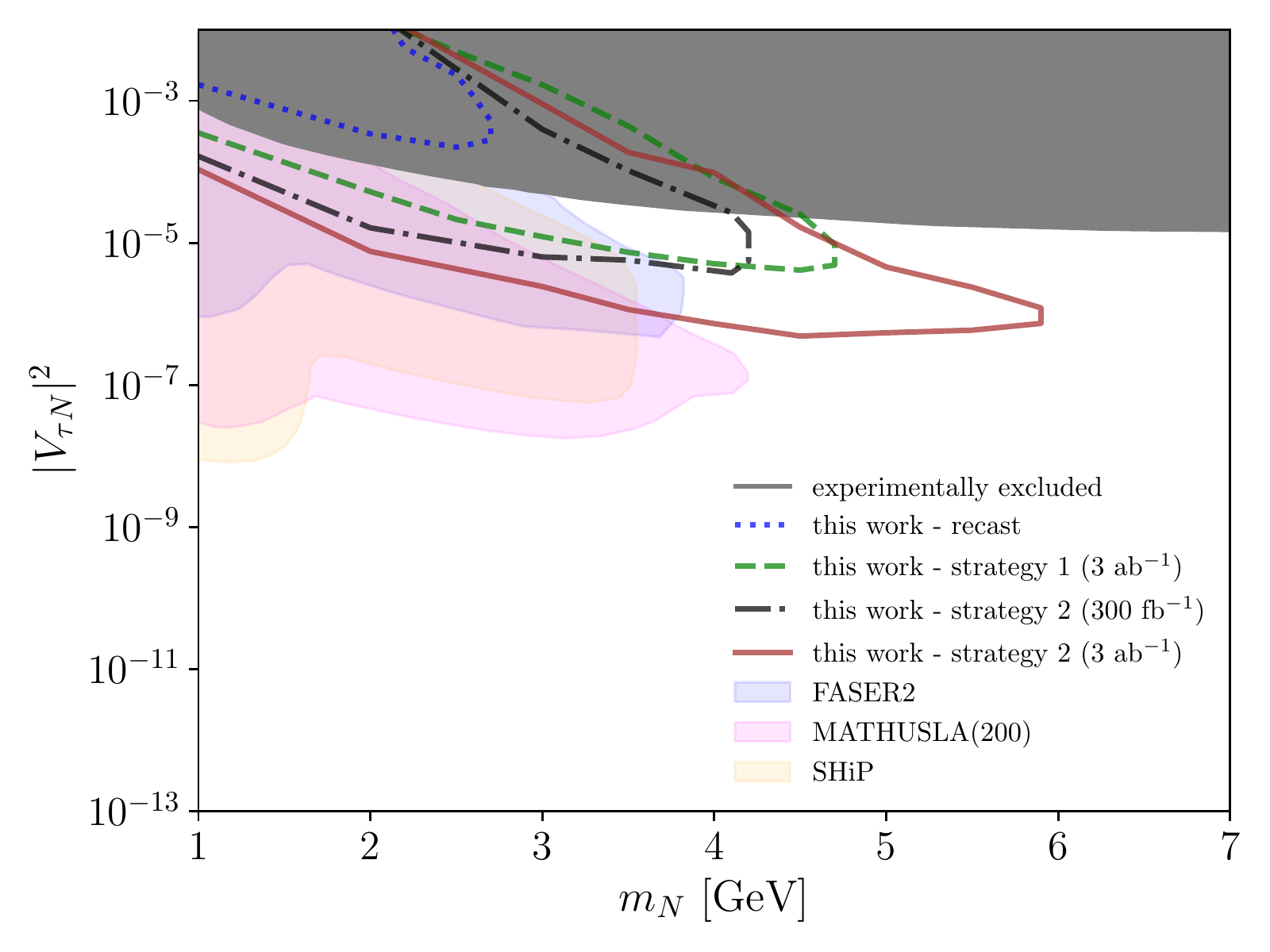}
    \includegraphics[width=0.49\textwidth]{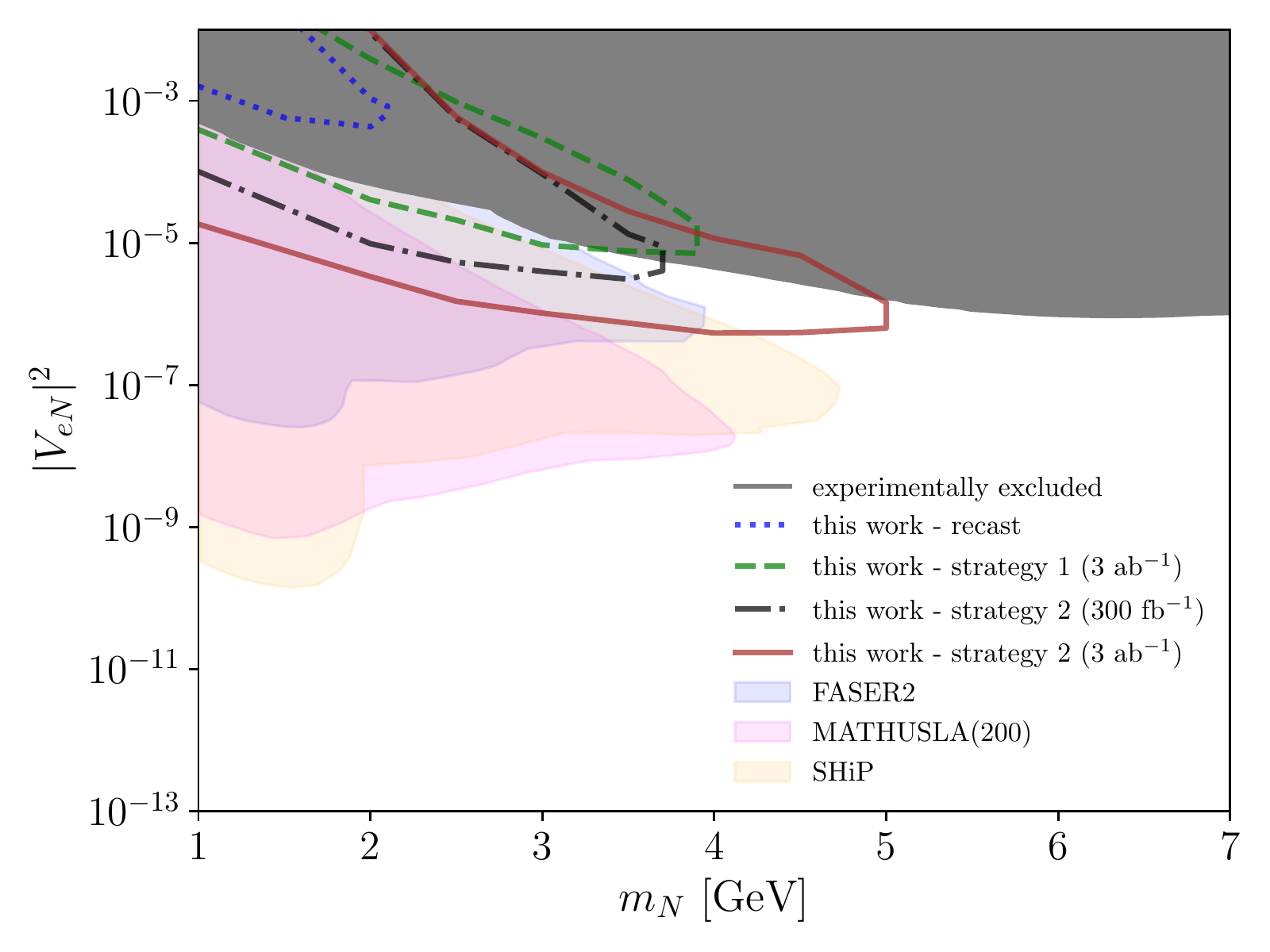}
    \caption{Projected sensitivity of the different proposed search strategies with a displaced shower signature in the CMS muon system. The minimal HNL scenario is considered with mixings in the $\tau$ and electron sectors, shown in the left and right panel, respectively.}
    \label{fig:limits}
\end{figure}

Figure~\ref{fig:limits} shows the estimated experimental sensitivity to HNLs with a displaced shower signatures at CMS. 
As explained in the previous section, the projected limits were calculated using two different trigger strategies and optimized for two datasets with different integrated luminosities. 
For the Run-2 dataset comprising of an integrated luminosity of 137~fb$^{-1}$, we performed a straightforward recast as described in section~\ref{sec:reinterpretation}, corresponding to the blue ``recast" contour in Figure~\ref{fig:limits}.

For strategy 1, which consists of an increased threshold requirement on the number of hits per cluster in order to reduce the expected background to near zero, as detailed in section~\ref{sec:reinterpretation}. 
This sensitivity estimate corresponds to a dataset with an integrated luminosity of 3 ab$^{-1}$ and which is represented by the dashed green ``strategy 1'' line in Figure~\ref{fig:limits}. 
For strategy 2, which uses a new dedicated displaced trigger, sensitivity estimates for datasets with luminosities of 300 fb$^{-1}$ and 3 ab$^{-1}$ are shown in black and brown, respectively.  

As can be seen in figure~\ref{fig:limits}, the sensitivities in $|V_{\tau N}|^2$ can reach values down to $|V_{\tau N}|^2 \sim 5\times 10^{-6}$ for $m_N \sim 5$ GeV with 3~ab$^{-1}$ of integrated luminosity using strategy 1. Upgrading to strategy 2 can improve the sensitivity in $|V_{\tau N}|^2$ down to $5\times 10^{-7}$ for the same integrated luminosity.  On the other hand, in the case of mixing with the electrons, the CMS muon system can reach values of the mixing parameter down to $|V_{e N}|^2 \sim  10^{-5}$ for $m_N \sim 4$ GeV using strategy 1 for an integrated luminosity of $3$ ab$^{-1}$. For the strategy 2 the limits can be improved up to    $|V_{e N}|^2 \sim  10^{-6}$ for $m_N \sim 5$ GeV for the same integrated luminosity.

Figure~\ref{fig:limits} also compares our limits with the current experimental bounds for this model, represented by the dark gray area at the top of each plot. These constraints refer to the limits from the DELPHI~\cite{Deppisch:2015qwa} and ATLAS experiments~\cite{ATLASnewHNL}. We also show for comparison the projected sensitivities from the proposed SHiP~\cite{SHiP:2015vad}, MATHUSLA~\cite{Helo:2018qej}, and FASER2~\cite{Kling:2018wct} experiments. As we can see, our forecasted limits can reach values of the mixing $|V_{\tau N}|^2$  two orders of magnitude smaller than current experimental bounds and are complementary to the proposed far detector experiments.

Finally, it is important to mention that in our analysis we have only considered $W$ boson decay as the main production mode of the HNLs.  However, for masses  $m_N \lesssim 5$ GeV, the HNLs can be also produced in meson decays or  tau lepton  decays if it's kinematically allowed.  
These contributions to the HNL production are expected to be important for the limits obtained using strategy 2, which has a new dedicated displaced trigger, and does not require triggering on high $\ptmiss$ nor a high $\pt$ prompt charged lepton.  
The analysis of the sensitivities of a displaced shower signature on HNLs coming from meson decays will be studied in a future work where we will also extend the range of our analysis for HNL masses smaller than 1 GeV.

\section{Summary} \label{sec:conclusion}  

The study of new search strategies for long-lived particle in the hunt for new physics is of critical importance to enhance the discovery potential of the LHC experiments. 
Such new phenomena can include an explanation for light neutrino masses in the SM, which motivates the study of models predicting long-lived heavy neutral leptons (HNLs).

In this work, we study a novel signature of a displaced particle shower resulting from the HNL decays that is reconstructed in the CMS muon detectors as a high-multiplicity hit cluster. 
As the muon detector shower signature has higher reconstruction efficiency for hadronic showers, we find this search is particularly sensitive to mixings in the $\tau-$sector, which is far less constrained experimentally than the scenario with mixing in the electron or muon sectors.  
We comment on theoretical motivations to search for an HNL coupled to taus, and highlight the inverse seesaw scenario, which can provide an HNL with tau mixing only.

We reinterpret a CMS search for long-lived neutral particles, and propose two new search strategies for HNLs optimized for the Run-3 and HL-LHC datasets. 
Our studies use a new Delphes module that encapsulates the CMS detector efficiency to the displaced signature decaying in the muon detectors. 
We provide sensitivity prospects using a new dedicated LLP trigger based on the detection of high hit-multiplicity muon detector showers that is being implemented in Run 3 of the LHC by CMS. 

We show that our strategies are sensitive to long-lived HNLs produced from the decays of $W$ bosons, and can access masses between 1 and 6 GeV, for values of the mixing as low as $|V_{\tau N}|^2 \sim 5\times 10^{-7}$. 
The projected sensitivity estimates cover gaps between existing LLP results and future dedicated LLP experiments, motivating further experimental studies at CMS.

\acknowledgments
We thank the Snowmass 2021 initiative for motivating this work. G.C. acknowledges support from ANID FONDECYT grant No. 11220237. G.C. and J.C.H. also acknowledge support from grant ANID FONDECYT grant No. 1201673 and ANID – Millennium Science Initiative Program ICN2019\_044. 
This work is supported by the Spanish grants PID2020-113775GB-I00 (AEI/10.13039/ 501100011033) and CIPROM/2021/054 (Generalitat Valenciana). 
We would like to thank the CMS Collaboration.   
CW and SX are partially supported by the U.S. Department of Energy, Office of Science, Office of High Energy Physics, under Award Number DE-SC0011925. 
This work  is part of CW's PhD thesis along with additional reinterpretation studies based solely on CMS published results and Delphes public codes \cite{CMS:2021juv,hepdata.104408.v2,delphes_pr}. 
CW, CP, and SX are grateful to the organizers and participants of the ``New ideas in detecting long-lived particles at the LHC'' workshop at LBNL in the Summer of 2018 where experimentalists and theorists gathered to generate new ideas on triggers and analysis strategies for long-lived particles searches at the LHC \cite{workshop} as well as the Fermilab LPC LLP group. 

\clearpage

\appendix
\section{Validation of the CMS search with new Delphes module}
\label{sec:validation}

The original model the CMS search in~\cite{CMS:2021juv} was interpreted in corresponds to a simplified twin Higgs model where a  SM Higgs boson $h$ decays to a pair of neutral long-lived scalars, $S$, each of which can then decay to $d$-quark pairs ($d\bar{d}$), $b$-quark pairs ($b\bar{b}$), and $\tau$ pairs ($\tau\bar{\tau}$). 

\begin{figure}[ht]
    \centering
    \includegraphics[width=0.45\textwidth]{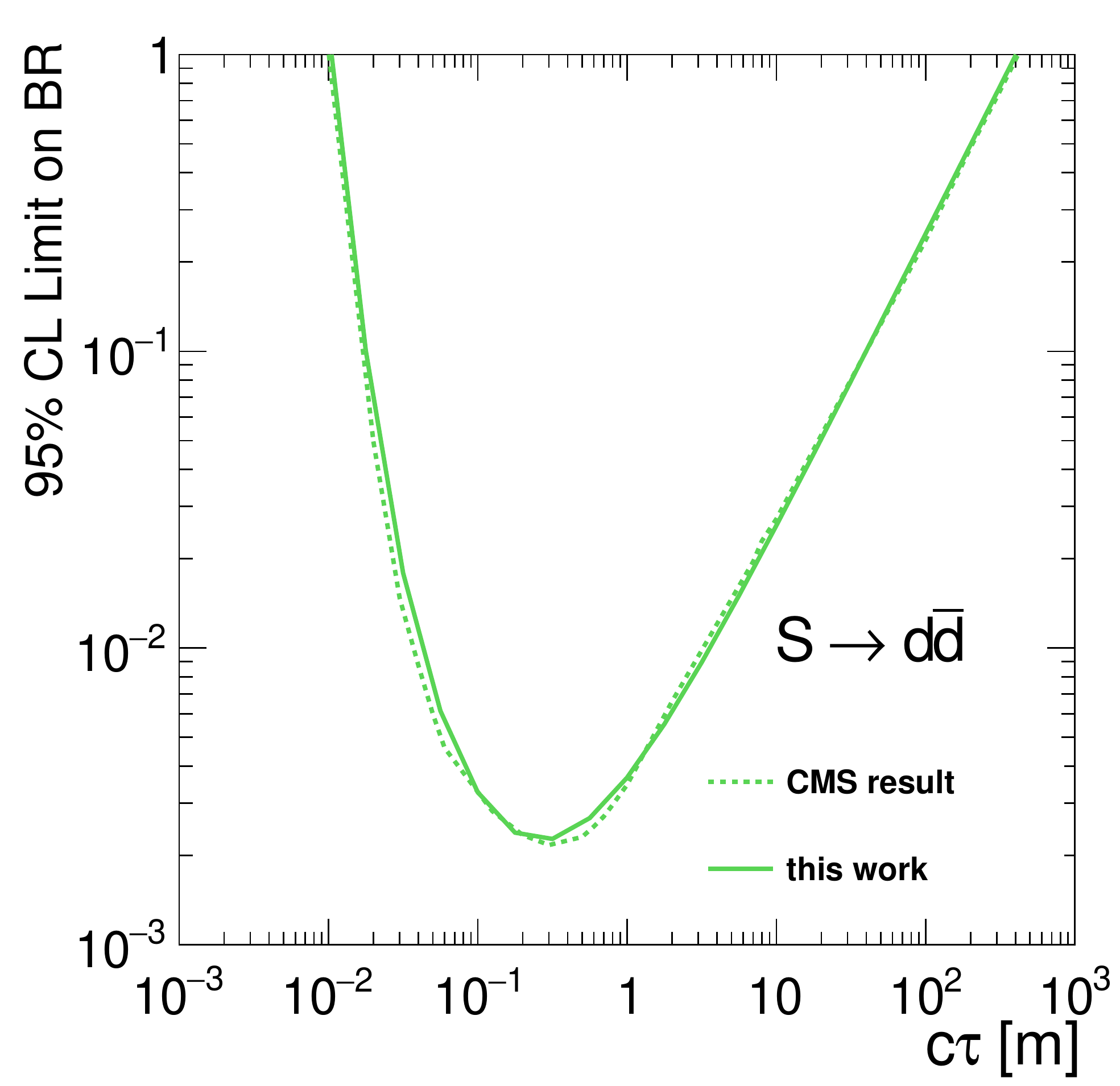}
    \includegraphics[width=0.45\textwidth]{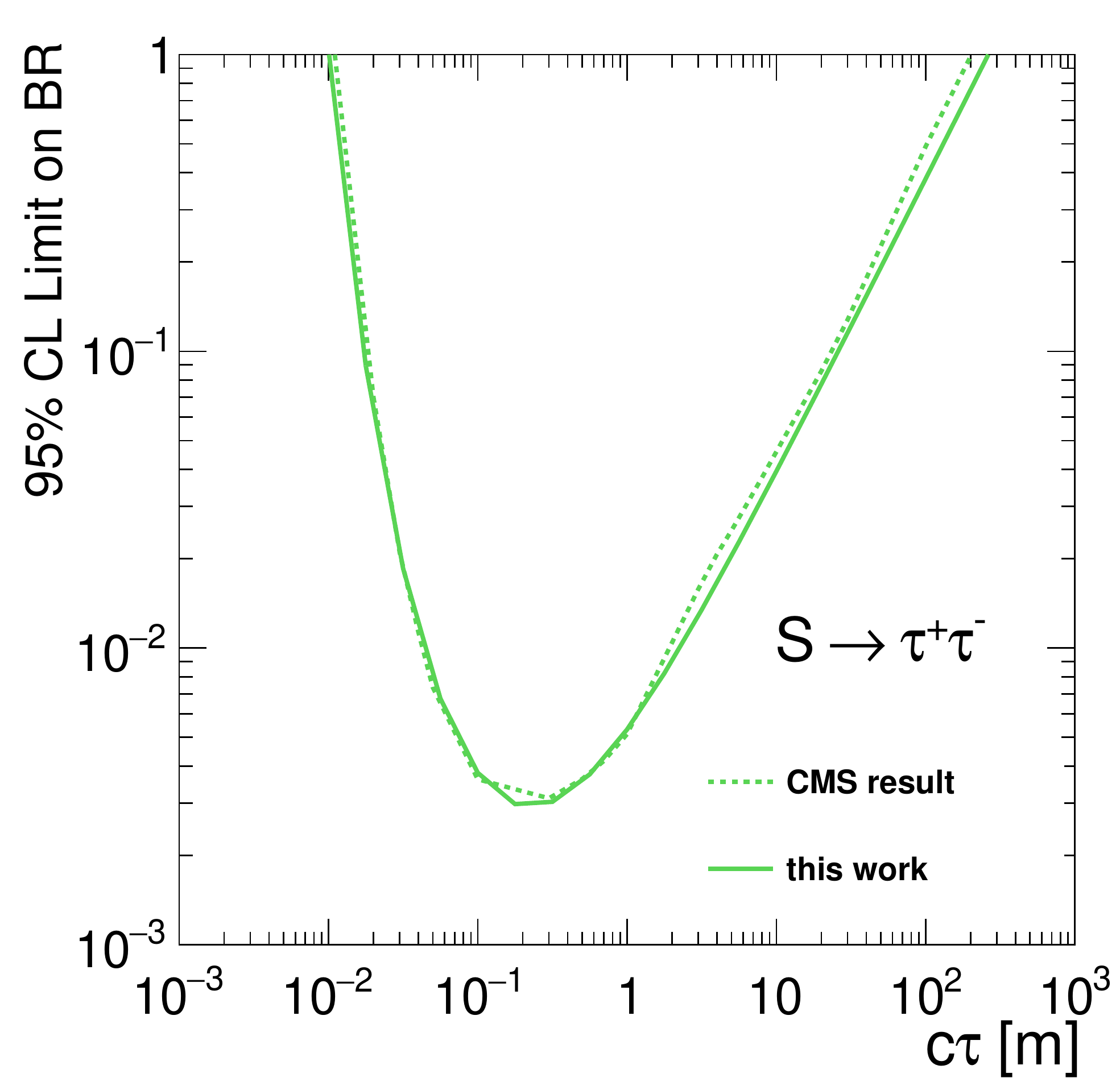}
    \caption{Comparison of the 95\% CL upper limits on the branching fraction ${\rm Br}(h\to SS)$ as functions of $c\tau$ derived with the standalone workflow (solid lines) and the CMS search (dashed lines). In deriving these limits we have considered a 7 GeV LLP decaying into $d$-quark pairs (left) and $\tau$ pairs (right). The limits from this work is shown to agree with CMS search to within $30\%$.}
    \label{fig:limit_validation}
\end{figure}

Given the signal and background yield estimates obtained when applying the reinterpretation procedure in section~\ref{sec:reinterpretation}, we evaluate the 95\% confidence level (CL) limits on the branching fraction ${\rm Br}(h\to SS)$. The observed $95\%$ CL upper limits on the branching fraction ${\rm Br}(h\to SS)$ for LLP scalar mass of $7$ GeV, as a function of $c\tau$, for decays $S\to d\bar d$ and $S\to\tau^{+}\tau^{-}$  are shown in Fig.~\ref{fig:limit_validation}. The limits evaluated using the detector response from Delphes agree with the CMS results within $30\%$ for all lifetimes evaluated.

\bibliography{refs}

\providecommand{\href}[2]{#2}\begingroup\raggedright\begin{thebibliography}{10}

\bibitem{Curtin:2018mvb}
D.~Curtin et~al., \emph{{Long-Lived Particles at the Energy Frontier: The
  MATHUSLA Physics Case}},
  \href{https://doi.org/10.1088/1361-6633/ab28d6}{\emph{Rept. Prog. Phys.}
  {\bfseries 82} (2019) 116201}
  [\href{https://arxiv.org/abs/1806.07396}{{\ttfamily 1806.07396}}].

\bibitem{Alimena:2019zri}
J.~Alimena et~al., \emph{{Searching for long-lived particles beyond the
  Standard Model at the Large Hadron Collider}},
  \href{https://doi.org/10.1088/1361-6471/ab4574}{\emph{J. Phys. G} {\bfseries
  47} (2020) 090501} [\href{https://arxiv.org/abs/1903.04497}{{\ttfamily
  1903.04497}}].

\bibitem{Minkowski:1977sc}
P.~Minkowski, \emph{{mu to e gamma at a Rate of One Out of 1-Billion Muon
  Decays?}},
  \href{https://doi.org/10.1016/0370-2693(77)90435-X}{\emph{Phys.Lett.}
  {\bfseries B67} (1977) 421}.

\bibitem{Yanagida:1979as}
T.~Yanagida, \emph{{Horizontal symmetry and masses of neutrinos}},
  {\emph{Conf.Proc.} {\bfseries C7902131} (1979) 95}.

\bibitem{GellMann:1980vs}
M.~Gell-Mann, P.~Ramond and R.~Slansky, \emph{{Complex Spinors and Unified
  Theories}}, {\emph{Conf. Proc.} {\bfseries C790927} (1979) 315}
  [\href{https://arxiv.org/abs/1306.4669}{{\ttfamily 1306.4669}}].

\bibitem{Mohapatra:1979ia}
R.N.~Mohapatra and G.~Senjanovic, \emph{{Neutrino Mass and Spontaneous Parity
  Violation}}, \href{https://doi.org/10.1103/PhysRevLett.44.912}{\emph{Phys.
  Rev. Lett.} {\bfseries 44} (1980) 912}.

\bibitem{Schechter:1980gr}
J.~Schechter and J.~Valle, \emph{{Neutrino Masses in SU(2) x U(1) Theories}},
  \href{https://doi.org/10.1103/PhysRevD.22.2227}{\emph{Phys. Rev.} {\bfseries
  D22} (1980) 2227}.

\bibitem{Helo:2013esa}
J.C.~Helo, M.~Hirsch and S.~Kovalenko, \emph{{Heavy neutrino searches at the
  LHC with displaced vertices}},
  \href{https://doi.org/10.1103/PhysRevD.89.073005}{\emph{Phys. Rev. D}
  {\bfseries 89} (2014) 073005}
  [\href{https://arxiv.org/abs/1312.2900}{{\ttfamily 1312.2900}}].

\bibitem{ATLASnewHNL}
{\scshape ATLAS} collaboration, \emph{{Search for heavy neutral leptons in
  decays of $W$ bosons using a dilepton displaced vertex in $\sqrt{s}=13$ TeV
  TeV pp collisions with the ATLAS detector }},  Tech. Rep. EXOT-2019-29
  (2022).

\bibitem{Aad:2019kiz}
{\scshape ATLAS} collaboration, \emph{{Search for heavy neutral leptons in
  decays of $W$ bosons produced in 13 TeV $pp$ collisions using prompt and
  displaced signatures with the ATLAS detector}},
  \href{https://doi.org/10.1007/JHEP10(2019)265}{\emph{JHEP} {\bfseries 10}
  (2019) 265} [\href{https://arxiv.org/abs/1905.09787}{{\ttfamily
  1905.09787}}].

\bibitem{CMS:2022fut}
{\scshape CMS} collaboration, \emph{{Search for long-lived heavy neutral
  leptons with displaced vertices in proton-proton collisions at $
  \sqrt{\mathrm{s}} $ =13 TeV}},
  \href{https://doi.org/10.1007/JHEP07(2022)081}{\emph{JHEP} {\bfseries 07}
  (2022) 081} [\href{https://arxiv.org/abs/2201.05578}{{\ttfamily
  2201.05578}}].

\bibitem{Cheung:2020buy}
K.~Cheung, Y.-L.~Chung, H.~Ishida and C.-T.~Lu, \emph{{Sensitivity reach on
  heavy neutral leptons and $\tau$-neutrino mixing $|U_{\tau N}|^2 $ at the
  HL-LHC}}, \href{https://doi.org/10.1103/PhysRevD.102.075038}{\emph{Phys. Rev.
  D} {\bfseries 102} (2020) 075038}
  [\href{https://arxiv.org/abs/2004.11537}{{\ttfamily 2004.11537}}].

\bibitem{Izaguirre:2015pga}
E.~Izaguirre and B.~Shuve, \emph{{Multilepton and Lepton Jet Probes of
  Sub-Weak-Scale Right-Handed Neutrinos}},
  \href{https://doi.org/10.1103/PhysRevD.91.093010}{\emph{Phys. Rev. D}
  {\bfseries 91} (2015) 093010}
  [\href{https://arxiv.org/abs/1504.02470}{{\ttfamily 1504.02470}}].

\bibitem{Beltran:2021hpq}
R.~Beltr\'an, G.~Cottin, J.C.~Helo, M.~Hirsch, A.~Titov and Z.S.~Wang,
  \emph{{Long-lived heavy neutral leptons at the LHC: four-fermion
  single-N$_{R}$ operators}},
  \href{https://doi.org/10.1007/JHEP01(2022)044}{\emph{JHEP} {\bfseries 01}
  (2022) 044} [\href{https://arxiv.org/abs/2110.15096}{{\ttfamily
  2110.15096}}].

\bibitem{Cottin:2018nms}
G.~Cottin, J.C.~Helo and M.~Hirsch, \emph{{Displaced vertices as probes of
  sterile neutrino mixing at the LHC}},
  \href{https://doi.org/10.1103/PhysRevD.98.035012}{\emph{Phys. Rev. D}
  {\bfseries 98} (2018) 035012}
  [\href{https://arxiv.org/abs/1806.05191}{{\ttfamily 1806.05191}}].

\bibitem{Abada:2018sfh}
A.~Abada, N.~Bernal, M.~Losada and X.~Marcano, \emph{{Inclusive Displaced
  Vertex Searches for Heavy Neutral Leptons at the LHC}},
  \href{https://doi.org/10.1007/JHEP01(2019)093}{\emph{JHEP} {\bfseries 01}
  (2019) 093} [\href{https://arxiv.org/abs/1807.10024}{{\ttfamily
  1807.10024}}].

\bibitem{Bondarenko:2019tss}
K.~Bondarenko, A.~Boyarsky, M.~Ovchynnikov, O.~Ruchayskiy and L.~Shchutska,
  \emph{{Probing new physics with displaced vertices: muon tracker at CMS}},
  \href{https://doi.org/10.1103/PhysRevD.100.075015}{\emph{Phys. Rev. D}
  {\bfseries 100} (2019) 075015}
  [\href{https://arxiv.org/abs/1903.11918}{{\ttfamily 1903.11918}}].

\bibitem{Drewes:2019fou}
M.~Drewes and J.~Hajer, \emph{{Heavy Neutrinos in displaced vertex searches at
  the LHC and HL-LHC}},
  \href{https://doi.org/10.1007/JHEP02(2020)070}{\emph{JHEP} {\bfseries 20}
  (2020) 070} [\href{https://arxiv.org/abs/1903.06100}{{\ttfamily
  1903.06100}}].

\bibitem{Boiarska:2019jcw}
I.~Boiarska, K.~Bondarenko, A.~Boyarsky, S.~Eijima, M.~Ovchynnikov,
  O.~Ruchayskiy et~al., \emph{{Probing baryon asymmetry of the Universe at LHC
  and SHiP}},  \href{https://arxiv.org/abs/1902.04535}{{\ttfamily 1902.04535}}.

\bibitem{Liu:2019ayx}
J.~Liu, Z.~Liu, L.-T.~Wang and X.-P.~Wang, \emph{{Seeking for sterile neutrinos
  with displaced leptons at the LHC}},
  \href{https://doi.org/10.1007/JHEP07(2019)159}{\emph{JHEP} {\bfseries 07}
  (2019) 159} [\href{https://arxiv.org/abs/1904.01020}{{\ttfamily
  1904.01020}}].

\bibitem{CMS:2021juv}
{\scshape CMS} collaboration, \emph{{Search for Long-Lived Particles Decaying
  in the CMS End Cap Muon Detectors in Proton-Proton Collisions at $\sqrt s$
  =13\,\,TeV}},
  \href{https://doi.org/10.1103/PhysRevLett.127.261804}{\emph{Phys. Rev. Lett.}
  {\bfseries 127} (2021) 261804}
  [\href{https://arxiv.org/abs/2107.04838}{{\ttfamily 2107.04838}}].

\bibitem{LLPCMSL1Trigger}
S.~Dildick, ``{Talk at Searching for long-lived particles at the LHC: Seventh
  workshop of the LHC LLP Community:
  \url{https://indico.cern.ch/event/863077/contributions/3850860/attachments/2045232/3427570/LLPWorkshop_20200523_SD.pdf}}.''
  May, 2020.

\bibitem{Alimena:2021mdu}
D.~Acosta et~al., \emph{{Review of opportunities for new long-lived particle
  triggers in Run 3 of the Large Hadron Collider}},
  \href{https://arxiv.org/abs/2110.14675}{{\ttfamily 2110.14675}}.

\bibitem{Mohapatra:1986bd}
R.~Mohapatra and J.~Valle, \emph{{Neutrino Mass and Baryon Number
  Nonconservation in Superstring Models}},
  \href{https://doi.org/10.1103/PhysRevD.34.1642}{\emph{Phys. Rev.} {\bfseries
  D34} (1986) 1642}.

\bibitem{delAguila:2008ir}
F.~del Aguila, S.~Bar-Shalom, A.~Soni and J.~Wudka, \emph{{Heavy Majorana
  Neutrinos in the Effective Lagrangian Description: Application to Hadron
  Colliders}},
  \href{https://doi.org/10.1016/j.physletb.2008.11.031}{\emph{Phys. Lett. B}
  {\bfseries 670} (2009) 399}
  [\href{https://arxiv.org/abs/0806.0876}{{\ttfamily 0806.0876}}].

\bibitem{Aparici:2009fh}
A.~Aparici, K.~Kim, A.~Santamaria and J.~Wudka, \emph{{Right-handed neutrino
  magnetic moments}},
  \href{https://doi.org/10.1103/PhysRevD.80.013010}{\emph{Phys. Rev. D}
  {\bfseries 80} (2009) 013010}
  [\href{https://arxiv.org/abs/0904.3244}{{\ttfamily 0904.3244}}].

\bibitem{Liao:2016qyd}
Y.~Liao and X.-D.~Ma, \emph{{Operators up to Dimension Seven in Standard Model
  Effective Field Theory Extended with Sterile Neutrinos}},
  \href{https://doi.org/10.1103/PhysRevD.96.015012}{\emph{Phys. Rev. D}
  {\bfseries 96} (2017) 015012}
  [\href{https://arxiv.org/abs/1612.04527}{{\ttfamily 1612.04527}}].

\bibitem{deSalas:2020pgw}
P.F.~de~Salas, D.V.~Forero, S.~Gariazzo, P.~Mart\'\i{}nez-Mirav\'e, O.~Mena,
  C.A.~Ternes et~al., \emph{{2020 global reassessment of the neutrino
  oscillation picture}},
  \href{https://doi.org/10.1007/JHEP02(2021)071}{\emph{JHEP} {\bfseries 02}
  (2021) 071} [\href{https://arxiv.org/abs/2006.11237}{{\ttfamily
  2006.11237}}].

\bibitem{Casas:2001sr}
J.A.~Casas and A.~Ibarra, \emph{{Oscillating neutrinos and $\mu \to e,
  \gamma$}}, \href{https://doi.org/10.1016/S0550-3213(01)00475-8}{\emph{Nucl.
  Phys. B} {\bfseries 618} (2001) 171}
  [\href{https://arxiv.org/abs/hep-ph/0103065}{{\ttfamily hep-ph/0103065}}].

\bibitem{Anamiati:2016uxp}
G.~Anamiati, M.~Hirsch and E.~Nardi, \emph{{Quasi-Dirac neutrinos at the LHC}},
  \href{https://doi.org/10.1007/JHEP10(2016)010}{\emph{JHEP} {\bfseries 10}
  (2016) 010} [\href{https://arxiv.org/abs/1607.05641}{{\ttfamily
  1607.05641}}].

\bibitem{Cordero-Carrion:2019qtu}
I.~Cordero-Carri\'on, M.~Hirsch and A.~Vicente, \emph{{General parametrization
  of Majorana neutrino mass models}},
  \href{https://doi.org/10.1103/PhysRevD.101.075032}{\emph{Phys. Rev. D}
  {\bfseries 101} (2020) 075032}
  [\href{https://arxiv.org/abs/1912.08858}{{\ttfamily 1912.08858}}].

\bibitem{Feng:2022inv}
J.L.~Feng et~al., \emph{{The Forward Physics Facility at the High-Luminosity
  LHC}},  in \emph{{2022 Snowmass Summer Study}}, 3, 2022
  [\href{https://arxiv.org/abs/2203.05090}{{\ttfamily 2203.05090}}].

\bibitem{Akhmedov:1995ip}
E.K.~Akhmedov, M.~Lindner, E.~Schnapka and J.~Valle, \emph{{Left-right symmetry
  breaking in NJL approach}},
  \href{https://doi.org/10.1016/0370-2693(95)01504-3}{\emph{Phys.Lett.}
  {\bfseries B368} (1996) 270}
  [\href{https://arxiv.org/abs/hep-ph/9507275}{{\ttfamily hep-ph/9507275}}].

\bibitem{Akhmedov:1995vm}
E.K.~Akhmedov, M.~Lindner, E.~Schnapka and J.~Valle, \emph{{Dynamical
  left-right symmetry breaking}},
  \href{https://doi.org/10.1103/PhysRevD.53.2752}{\emph{Phys.Rev.} {\bfseries
  D53} (1996) 2752} [\href{https://arxiv.org/abs/hep-ph/9509255}{{\ttfamily
  hep-ph/9509255}}].

\bibitem{Alwall:2011uj}
J.~Alwall, M.~Herquet, F.~Maltoni, O.~Mattelaer and T.~Stelzer, \emph{{MadGraph
  5 : Going Beyond}},
  \href{https://doi.org/10.1007/JHEP06(2011)128}{\emph{JHEP} {\bfseries 06}
  (2011) 128} [\href{https://arxiv.org/abs/1106.0522}{{\ttfamily 1106.0522}}].

\bibitem{Alwall:2014hca}
J.~Alwall, R.~Frederix, S.~Frixione, V.~Hirschi, F.~Maltoni, O.~Mattelaer
  et~al., \emph{{The automated computation of tree-level and next-to-leading
  order differential cross sections, and their matching to parton shower
  simulations}}, \href{https://doi.org/10.1007/JHEP07(2014)079}{\emph{JHEP}
  {\bfseries 07} (2014) 079} [\href{https://arxiv.org/abs/1405.0301}{{\ttfamily
  1405.0301}}].

\bibitem{Sjostrand:2014zea}
T.~Sj\"ostrand, S.~Ask, J.R.~Christiansen, R.~Corke, N.~Desai, P.~Ilten et~al.,
  \emph{{An introduction to PYTHIA 8.2}},
  \href{https://doi.org/10.1016/j.cpc.2015.01.024}{\emph{Comput. Phys. Commun.}
  {\bfseries 191} (2015) 159}
  [\href{https://arxiv.org/abs/1410.3012}{{\ttfamily 1410.3012}}].

\bibitem{delphes}
J.~de~Favereau, C.~Delaere, P.~Demin, A.~Giammanco, V.~Lemaître, A.~Mertens
  et~al., \emph{Delphes 3: a modular framework for fast simulation of a generic
  collider experiment},
  \href{https://doi.org/10.1007/jhep02(2014)057}{\emph{Journal of High Energy
  Physics} {\bfseries 2014} (2014) }.

\bibitem{delphes_pr}
C.~Wang, ``{Dedicated Delphes Module:
  \url{https://github.com/delphes/delphes/pull/103}}.'' March, 2022.

\bibitem{Degrande:2016aje}
C.~Degrande, O.~Mattelaer, R.~Ruiz and J.~Turner, \emph{{Fully-Automated
  Precision Predictions for Heavy Neutrino Production Mechanisms at Hadron
  Colliders}}, \href{https://doi.org/10.1103/PhysRevD.94.053002}{\emph{Phys.
  Rev. D} {\bfseries 94} (2016) 053002}
  [\href{https://arxiv.org/abs/1602.06957}{{\ttfamily 1602.06957}}].

\bibitem{Mangano:2006rw}
M.L.~Mangano, M.~Moretti, F.~Piccinini and M.~Treccani, \emph{{Matching matrix
  elements and shower evolution for top-quark production in hadronic
  collisions}},
  \href{https://doi.org/10.1088/1126-6708/2007/01/013}{\emph{JHEP} {\bfseries
  01} (2007) 013} [\href{https://arxiv.org/abs/hep-ph/0611129}{{\ttfamily
  hep-ph/0611129}}].

\bibitem{dyturbo}
S.~Camarda, M.~Boonekamp, G.~Bozzi, S.~Catani, L.~Cieri, J.~Cuth et~al.,
  \emph{Dyturbo: fast predictions for drell–yan processes},
  \href{https://doi.org/10.1140/epjc/s10052-020-7757-5}{\emph{The European
  Physical Journal C} {\bfseries 80} (2020) }.

\bibitem{Mertens:2015kba}
A.~Mertens, \emph{{New features in Delphes 3}},
  \href{https://doi.org/10.1088/1742-6596/608/1/012045}{\emph{J. Phys. Conf.
  Ser.} {\bfseries 608} (2015) 012045}.

\bibitem{hepdata.104408.v2}
{CMS Collaboration}, ``{Search for long-lived particles decaying in the CMS
  endcap muon detectors in proton-proton collisions at $\sqrt{s} = $ 13 TeV
  (Version 2)}.'' {\href{https://doi.org/10.17182/hepdata.104408.v2}{HEPData
  (collection)}}, 2021.

\bibitem{Apollinari:2015wtw}
G.~Apollinari, O.~Br\"uning, T.~Nakamoto and L.~Rossi, \emph{{High Luminosity
  Large Hadron Collider HL-LHC}},
  \href{https://doi.org/10.5170/CERN-2015-005.1}{\emph{CERN Yellow Rep.} (2015)
  1} [\href{https://arxiv.org/abs/1705.08830}{{\ttfamily 1705.08830}}].

\bibitem{CMS:2667167}
{\scshape CMS} collaboration, \emph{{A MIP Timing Detector for the CMS Phase-2
  Upgrade}},  Tech. Rep.
  \href{https://cds.cern.ch/record/2667167}{CERN-LHCC-2019-003, CMS-TDR-020},
  CERN, Geneva (Mar, 2019).

\bibitem{Read_2002}
A.L.~Read, \emph{{Presentation of search results: The CL(s) technique}},
  \href{https://doi.org/10.1088/0954-3899/28/10/313}{\emph{J. Phys. G}
  {\bfseries 28} (2002) 2693}.

\bibitem{Deppisch:2015qwa}
F.F.~Deppisch, P.S.~Bhupal~Dev and A.~Pilaftsis, \emph{{Neutrinos and Collider
  Physics}}, \href{https://doi.org/10.1088/1367-2630/17/7/075019}{\emph{New J.
  Phys.} {\bfseries 17} (2015) 075019}
  [\href{https://arxiv.org/abs/1502.06541}{{\ttfamily 1502.06541}}].

\bibitem{SHiP:2015vad}
{\scshape SHiP} collaboration, \emph{{A facility to Search for Hidden Particles
  (SHiP) at the CERN SPS}},  \href{https://arxiv.org/abs/1504.04956}{{\ttfamily
  1504.04956}}.

\bibitem{Helo:2018qej}
J.C.~Helo, M.~Hirsch and Z.S.~Wang, \emph{{Heavy neutral fermions at the
  high-luminosity LHC}},
  \href{https://doi.org/10.1007/JHEP07(2018)056}{\emph{JHEP} {\bfseries 07}
  (2018) 056} [\href{https://arxiv.org/abs/1803.02212}{{\ttfamily
  1803.02212}}].

\bibitem{Kling:2018wct}
F.~Kling and S.~Trojanowski, \emph{{Heavy Neutral Leptons at FASER}},
  \href{https://doi.org/10.1103/PhysRevD.97.095016}{\emph{Phys. Rev. D}
  {\bfseries 97} (2018) 095016}
  [\href{https://arxiv.org/abs/1801.08947}{{\ttfamily 1801.08947}}].

\bibitem{workshop}
``"{New} ideas in detecting long-lived particles at the {LHC Workshop}".''
  \url{https://indico.physics.lbl.gov/event/633/}.

\end{thebibliography}\endgroup
\bibliographystyle{jhep}
\end{document}